\documentclass[9pt,twocolumn,twoside]{opticajnl}
\journal{opticajournal} 
\usepackage[T1]{fontenc}
\usepackage{svg}
\usepackage{amsmath} 
\usepackage{relsize}
\setboolean{shortarticle}{true}
\usepackage{lineno}

\title{Efficient multiphoton microscopy with picosecond laser pulses}

\author[ ]{Katarzyna Kunio}
\author[ *]{Jakub Bogusławski}
\author[ ]{Grzegorz Soboń}

\affil[ ]{Laser \& Fiber Electronics Group, Faculty of Electronics, Photonics and Microsystems, Wrocław University of Science and Technology, \newline Wybrzeże Wyspiańskiego 27, 50-370 Wrocław, Poland}

\affil[*]{jakub.boguslawski@pwr.edu.pl}

\begin{abstract}
Multiphoton microscopes employ femtosecond lasers as light sources because the high peak power of the ultrashort pulse allows for multiphoton excitation of fluorescence in the examined sample. However, such short pulses are susceptible to broadening in a microscope’s highly dispersive optical elements and require careful dispersion management, otherwise decreasing excitation efficiency. Here, we have developed a 10 nJ Yb:fiber picosecond laser with an integrated pulse picker unit and evaluated its performance in multiphoton microscopy. Our results show that performance comparable to femtosecond pulses can be obtained with picosecond pulses only by reducing the pulse repetition rate and that such pulses are significantly less prone to the effect of chromatic dispersion. These findings proved that the temporal compression of fiber lasers is not always efficient, and it can be omitted by using a smaller and easier-to-use all-fiber setup. 
\end{abstract}

\begin{document}

\maketitle
Multiphoton microscopy \cite{Hoover-1, Denk-2} enables numerous applications, such as noninvasive imaging of single cells in living tissues \cite{Pillai-3, Tsakanova-4}. It has many advantages compared to its predecessor, single-photon (confocal) microscopy, e.g., deeper tissue penetration and reduced damage caused by direct ultraviolet light \cite{Konig-5}. Multiphoton excitation is possible using femtosecond lasers operating in the near-infrared range \cite{Rohrbacher-6} due to their high peak power and short pulse duration. Longer wavelengths contribute to reduced tissue scattering, allowing for greater penetration depth. Moreover, the short pulse duration minimizes the time during which the specimen is illuminated by the beam, reducing the photodamage of the sample.
\vspace{-0.85pt}

Two-photon excitation (TPE) is the most prominent modality of multiphoton microscopy, and one of the most popular laser sources for this application are Ti:sapphire lasers \cite{Konig-5, Rohrbacher-6, Paoli-7}. They offer a broad wavelength tuning, short pulse durations, and the ability to generate high peak powers needed for inducting two-photon absorption. However, they are also complex, costly, and require specialized maintenance, stopping multiphoton microscopy from becoming widely used. Fiber lasers have emerged as an attractive alternative, covering a broad spectrum of wavelengths using Nd, Yb, and frequency-doubled Er-doped lasers \cite{Kim-8, Chen-9, Sun-10, Stachowiak-11}. They offer a smaller footprint and easier cooling in an all-fiber, maintenance-free format. However, Nd and Yb:fiber lasers usually generate chirped, picosecond pulses and require bulk-optics compressors outside the laser cavity \cite{Szczepanek-12}.

The high peak power of femtosecond lasers used in multiphoton microscopy increases the likelihood of two-photon absorption, as the probability of two-photon absorption is proportional to the square of the instantaneous intensity. Regardless of the laser type, an ultrashort pulse propagating through highly dispersive optical elements of a microscope, such as scan and tube lenses or an objective \cite{Radzewicz-13, Busing-14}, will lead to broadening. This effect is also very challenging for multiphoton endoscopes \cite{Jung-15}. The shorter the pulse, the more significant the effect, which, in turn, decreases the excitation efficiency. Most commercially available lasers offer compressed pulses directly at the laser output. Efficient multiphoton imaging often requires additional dispersion pre-compensation \cite{Horton-16, Perez-17, Boguslawski-18}. Tremendous efforts have been made to deliver a perfectly compressed pulse directly to the sample plane, including advanced methods, such as multiphoton intrapulse interference phase scan (MIIPS) \cite{Liu-19, Murashova-20, Palczewska-21}. In practice, compensating for the microscope’s dispersion is expensive and challenging for end users without a technical background, such as biologists or medical professionals.

\begin{figure}[t]
\vspace{-15pt}
\centering
\includegraphics[width=\linewidth]{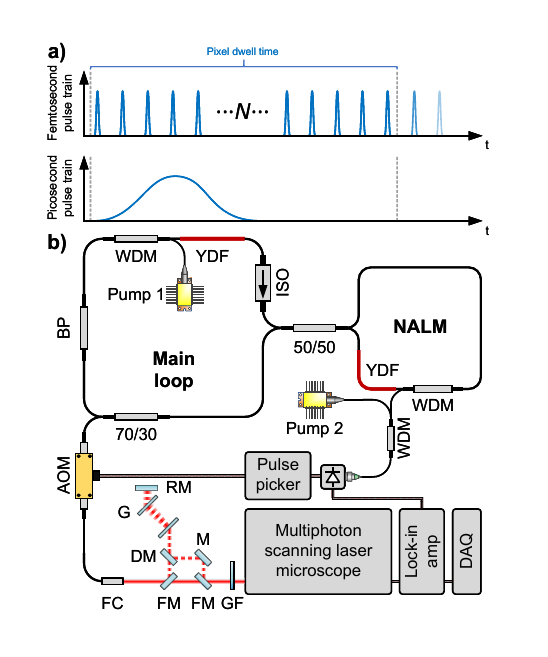}
\vspace{-35pt}
\caption{Principle of operation and experimental layout. a) Femtosecond and picosecond pulse trains with the same duty cycle and the same peak power. b) Experimental setup of picosecond fiber laser and multiphoton microscope system. BP – bandpass filter, YDF – Yb-doped fiber, WDM – wavelength-division multiplier, ISO – isolator, AOM – acousto-optic modulator, FC – fiber collimator, FM – flip mirror, DM – D-shaped mirror, G – diffraction grating, RM- retro mirror, M – mirror, GF – gradient index filter, DAQ – data acquisition card.}
\label{fig1}
\vspace{-15pt}
\end{figure}

It is often thought that femtosecond pulses are necessary for the TPE of fluorescent molecules. However, it has been proven that the same effect can be achieved by using a longer pulse with a lower repetition rate and the same peak power (\textit{$p_{peak}$}) \cite{Karpf-22}. In TPE, the fluorescence signal depends on the pulse train’s duty cycle, which is a product of pulse duration (\textit{$\tau_p$}) and pulse repetition rate ($f_{rep}$) \cite{Song-23}:
\begin{equation}
    n \sim \frac{p^2_{avr}}{\tau_p \cdot f_{rep}} = p_{peak}^2\cdot \tau_p \cdot f_{rep},
\label{eq1}
\end{equation}
where \textit{n} is the average number of photons that the fluorescing medium emits per second, \textit{$p_{avr}$} is the average power. The same effect typically achieved by reducing the pulse duration can be accomplished by longer pulses with reduced repetition rate, as long as they have the same \textit{$p_{avr}$} (and hence duty cycle). The longer pulse can be considered an equivalent of multiple shorter pulses stacked together, as shown in Fig. \ref{fig1}(a). Practically, the \textit{$f_{rep}$} can be controlled by the pulse picker unit outside the laser cavity. 

Here, we experimentally show that an efficient TPE is possible by directly using picosecond pulses with reduced \textit{$f_{rep}$}. We designed a monolithic, all-polarization-maintaining Yb:fiber laser producing 10 ps pulses with $\sim$10 nJ energy directly at the oscillator output. The laser was connected with a fiber-coupled pulse picker unit and directly applied in multiphoton microscopy with reduced \textit{$f_{rep}$} (we refer to this case as a picosecond pulse train). We compared those results with images obtained using a compressed 205 fs pulse train without reducing the \textit{$f_{rep}$}, hence keeping the same duty cycle (we refer to this case as femtosecond pulse train). We show the advantage of utilizing the picosecond pulse train with pulse picking in two imaging modalities: TPE fluorescence and second harmonic generation (SHG) microscopy. We demonstrated that picosecond excitation is less susceptible to chromatic dispersion (of a microscope or other optics), thus making it a preferred solution for endoscopic applications. The omission of the pulse compression unit enables the all-fiber configuration and lowers the setup costs and complexity.

A schematic diagram of the laser system is illustrated in Fig. \ref{fig1}(b). Pulses were first generated by a self-starting Yb:fiber figure-eight laser consisting of two cavity sections: the main loop and the nonlinear amplification loop mirror (NALM) \cite{Chernysheva-24}, each pumped with a 976 nm laser diode followed by an isolator and a pump protector. The NALM loop comprised a reflection-type wavelength-division multiplexer (WDM) and an asymmetrically placed active Yb-doped fiber (YDF, Coherent PM-YSF-HI-HP). The main loop comprised a YDF, an isolator (ISO), a WDM, a 10-nm bandpass filter (BP), and a fiber coupler, which led 70\% of the optical power out of the system. All fibers and components were polarization-maintaining. The cavity length is 13.6 m, corresponding to the fundamental repetition rate of 15.2 MHz. The equipment used for laser characterization is listed in the Supplemental Document.

To change the \textit{$f_{rep}$}, the pulse picker unit consisting of an acousto-optic modulator (AOM, G\&H Fiber-Q) and an electronic driver (AA Optoelectronic PPKS200) was added to the setup. The modulator can pick out single pulses at a desired \textit{$f_{rep}$}. The transmission of the pulses was synchronized with the original pulse train supplied to the photodiode (Thorlabs PDA05CF2) from the previously unused port in the WDM. The user can change the \textit{$f_{rep}$} of the pulse train by dividing its value by a positive integer. The output beam was collimated (FC, Schäfter+Kirchhoff 60FC-4-A11) and was directed to the custom-built multiphoton scanning laser microscope (described in \cite{Boguslawski2-25}), working in TPE epi-fluorescence or SHG mode. The photomultiplier tube output (Thorlabs PMT2101) was connected to a lock-in amplifier (Zurich Instruments HF2LI) with a reference frequency from the photodiode. The demodulated signal was sampled with a data acquisition card (DAQ, NI PCIe-6363). The dwell time of one pixel was set to 10 µs, and the size of one image was 1024×1024 pixels; in all cases, 50 frames were averaged.

The output laser beam could be directed through a typical Treacy compressor \cite{Treacy-26} using mirrors on a flip mount. It comprised two parallel transmission diffraction gratings (Coherent LightSmyth, 1000 grooves/mm, G) with an estimated group delay dispersion of -1.17 ps$^2$ and transmission of 79\%. This way, we could form two types of excitation pulse trains: i) a picosecond pulse train with adjustable \textit{$f_{rep}$}, and ii) a femtosecond pulse train where, for comparison, we keep the \textit{$f_{rep}$} unchanged with respect to the repetition rate of the oscillator. Many commercially available lasers do not have a built-in dispersion precompensation for microscope’s optics, and end users often do not have the technical capabilities to build it. To reflect this scenario, we did not pre-compensate the microscope’s chromatic dispersion, which also allows us to investigate differences in the effect of dispersion on pico- and femtosecond pulses. Both beams passed through the gradient filter (GF, Thorlabs NDL-10C-4) on their way into the multiphoton microscope, allowing the adjustment of the average power. This allows us to keep the average powers of pico- and femtosecond pulse trains the same for a fair comparison.
\begin{figure}[b]
\vspace{-15pt}
\centering
\includegraphics[width=\linewidth]{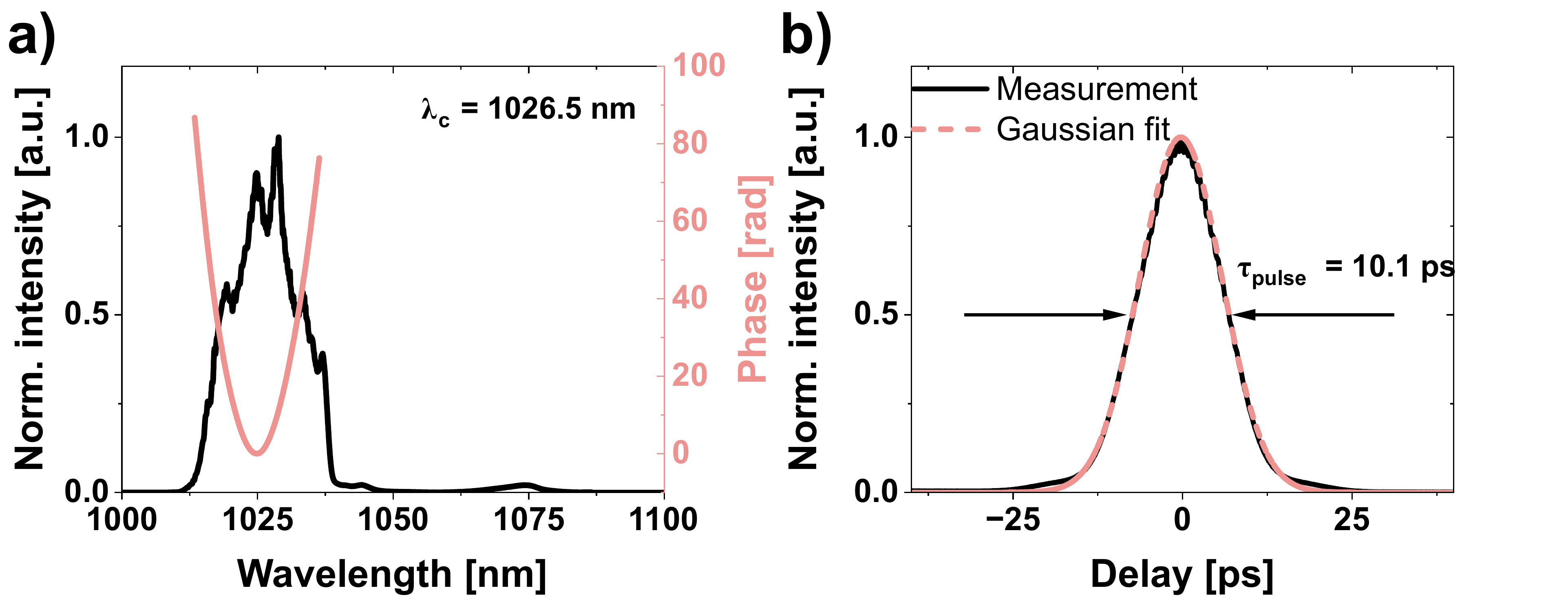}
\caption{Characterization of the laser directly at the oscillator output. a) The optical spectrum with spectral phase (red). b) Autocorrelation of the pulse and the Gaussian fit.}
\label{fig2}
\vspace{-15pt}
\end{figure}

Figure \ref{fig2} summarizes the laser performance directly at the oscillator output. The oscillator generated pulses at the center wavelength of 1026.5 nm with full width at half-maximum (FWHM) of 15.6 nm [Fig. \ref{fig2}(a)]. The spectrum also consisted of a smaller peak at around 1075 nm (shift of 440 cm$^{-1}$) caused by Raman scattering. We established numerically and experimentally that the peak holds only 1.5\% of the average power. The pulse duration was 10.1 ps, assuming a Gaussian pulse shape, as shown in Fig. \ref{fig2}(b). The output power of 148.5 mW (resulting in the pulse energy of $\sim$10 nJ) is comparable with state-of-the-art high-power oscillators \cite{Deng-27}. It ensures enough power is left for the imaging after adding more elements, and excellent power stability guarantees the same average power at each pixel of the image. The power stability and noise performance are included in section 1 of the Supplemental Document.

Figure \ref{fig3} summarizes the temporal properties of pico- and femtosecond pulse trains at the microscope entrance and after propagating through the microscope’s optics. After propagating through the AOM, pulse duration [Fig. \ref{fig3}(a)] got slightly longer (10.30 ps) because of propagation in approx. 1.25 m of added fiber. Adding the AOM caused a loss of $\sim$40\%, resulting in a pulse energy of 5.6 nJ; however, there was no significant change in the optical spectrum [see Fig. S4(a)]. Figure \ref{fig3}(b) shows the temporal profile of the femtosecond pulse after compression measured by spectral phase interferometry for direct electric-field reconstruction (SPIDER) technique. We obtained a FWHM duration of 205 fs with approximately 50\% of the peak power of the transform-limited pulse. The optical spectrum was centered at 1029 nm, as shown in Fig. S4(b). After propagating through the microscope (i.e., in the sample plane), the picosecond pulse duration remained very similar, while the femtosecond pulse got stretched to 227 fs (FWHM) with approximately 35\% of the peak power of the transform-limited pulse. To further evaluate the effect of chromatic dispersion on both pulse trains, we measured pulse widths at the sample plane with added glass blocks of various dispersion (within 1880 – 4400 fs$^2$, reflecting the case in which additional optical elements are added to the system or microscope objective is exchanged). The results shown in Fig. S5 and S6 demonstrate that picosecond pulses are significantly less susceptible to the effect of dispersion.

\begin{figure}[b]
\vspace{-15pt}
\centering
\includegraphics[width=\linewidth]{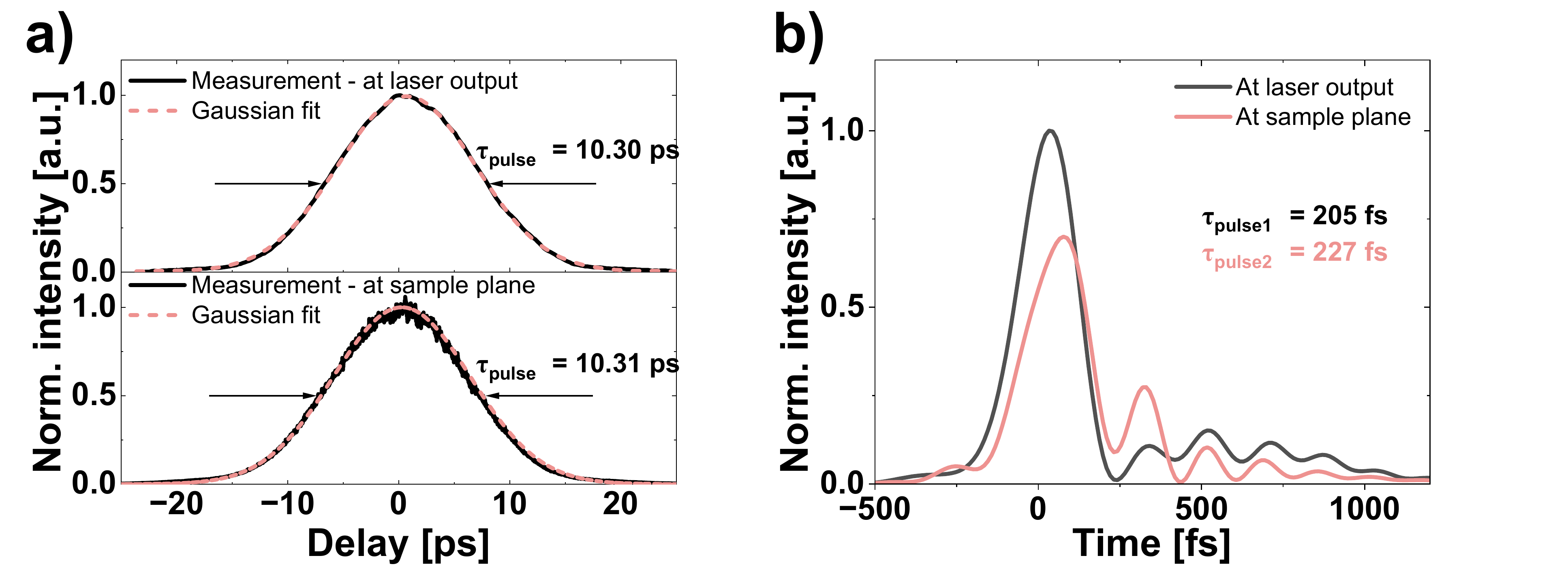}
\caption{Effect of microscope’s dispersion on pico- and femtosecond pulses. a) Comparison of the autocorrelation of the picosecond pulse with a Gaussian fit at the laser output (top) and in the sample plane (bottom), and b) comparison of the reconstructed temporal profile of the femtosecond pulse at the laser output and in the sample plane.}
\label{fig3}
\vspace{-15pt}
\end{figure}

\begin{figure}[t]
\centering
\includegraphics[width=\linewidth]{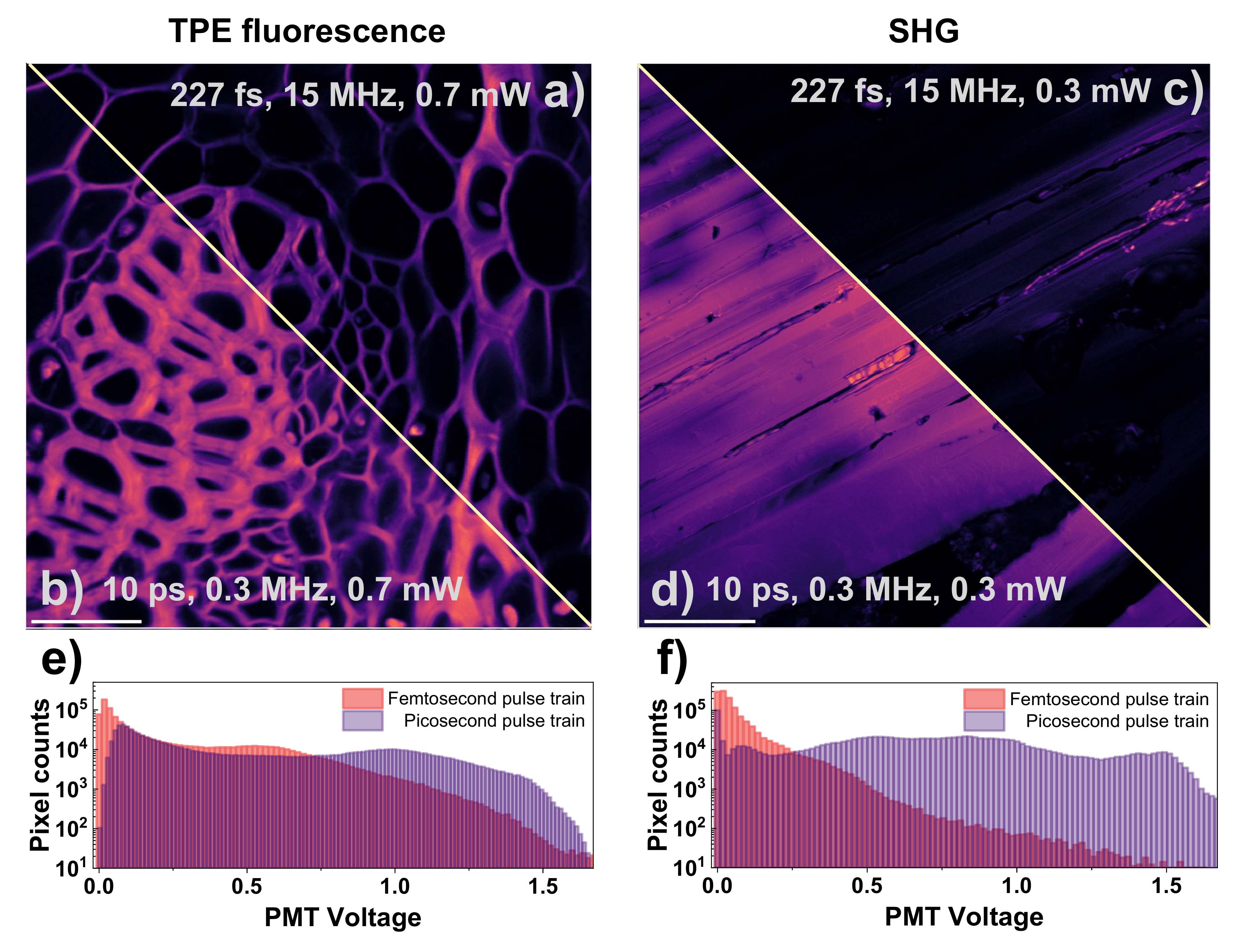}
\caption{Comparison of imaging with pico- and femtosecond pulse trains with the same average power and comparable duty cycle. TPE fluorescence images of \textit{convallaria majalis} sample obtained using: (a) femtosecond and (b) picosecond pulse trains (scale bar: 75 $\mu$m). SHG images of urea microcrystals using (c) femtosecond and (d) picosecond pulse trains (scale bar: 200 $\mu$m). Histograms of pixel intensity for both pulse trains for (e) TPE fluorescence and (f) SHG.}
\label{fig4}
\vspace{-15pt}
\end{figure}

\begin{figure}[h!b]
\vspace{-15pt}
\centering
\includegraphics[width=\linewidth]{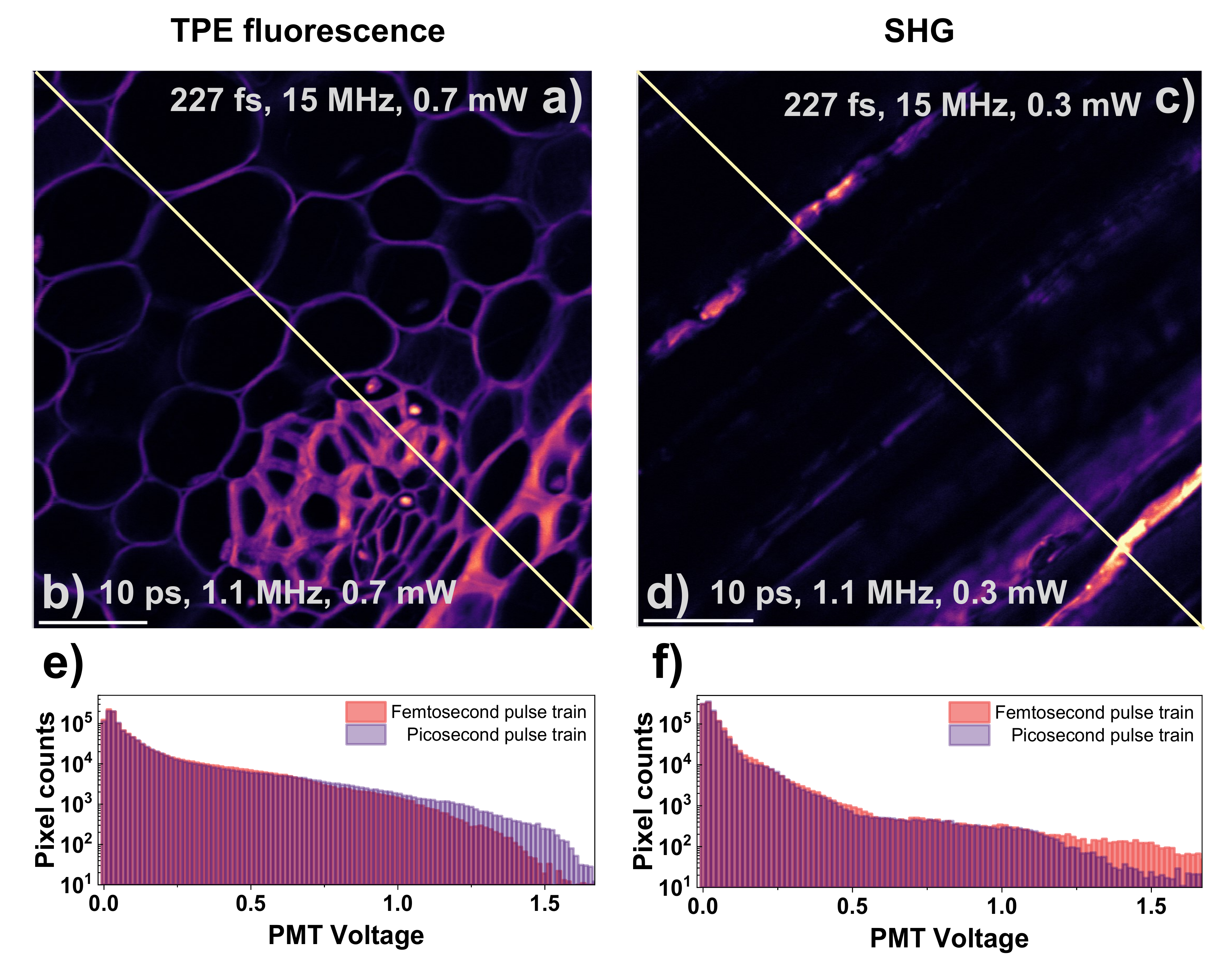}
\caption{Comparison of imaging with pico- and femtosecond pulse trains with the same average power and comparable PMT signal. TPE fluorescence images of \textit{convallaria majalis} sample obtained using: (a) femtosecond and (b) picosecond pulse trains (scale bar: 75 $\mu$m). SHG images of urea microcrystals using (c) femtosecond and (d) picosecond pulse trains (scale bar: 200 $\mu$m). Histograms of pixel intensity for both pulse trains for (e) TPE fluorescence and (f) SHG.}
\label{fig5}
\vspace{-15pt}
\end{figure}

Next, we evaluated the performance of picosecond vs. femtosecond pulses in multiphoton microscopy. First, we compare the femto- and picosecond pulse trains with the same average power and comparable duty cycle. To match the duty cycle of the femtosecond pulse train, i.e., 227 fs $\cdot$ 15.2 MHz = 3.4 $\times10^{-6}$, the \textit{$f_{rep}$} of the picosecond pulse train was reduced to 0.3 MHz (Table \ref{tab1} lists all relevant parameters). Figures \ref{fig4}(a)-(b) show the results of imaging \textit{convallaria majalis} root transverse section stained with an acridine orange, a nucleic acid-specific fluorophore; the average powers of both pulse trains were equal to 0.7 mW at the sample. Figures \ref{fig4}(c) and (d) show the results of SHG imaging a urea microcrystals with 0.3 mW average power. The images received using the picosecond pulse train were brighter and, thus, had better contrast. Figures \ref{fig4}(e)-(f) compare pixel intensity histograms and show a significant shift toward higher voltages for the picosecond pulse train. Using a picosecond pulse train for TPE fluorescence resulted in a 54\% higher fluorescence signal, while the SHG image had a 90\% higher signal. We point out that Eq. \ref{eq1} does not consider the actual pulse shape but only the FWHM value. While the picosecond pulse autocorrelation is perfectly fitted with the Gaussian function, the pulse after the compression and propagation through the microscope’s optics has post-pulses and, consequently, lower peak power than the picosecond pulse. Following Eq. \ref{eq1}, reduced peak power can be balanced by increasing the \textit{$f_{rep}$}. We gradually increased the \textit{$f_{rep}$} of the picosecond pulse train while adjusting its average power using the gradient index filter to obtain the same image quality for both cases. Figures \ref{fig5}(a)-(d) demonstrate images obtained with 1.1 MHz \textit{$f_{rep}$} and their respective histograms [Fig. \ref{fig5}(e)-(f)].

\begin{table}[t!]
\vspace{-15pt}
\caption{Comparison of parameters of femtosecond and picosecond pulse trains used in the experiment.}
\resizebox{\columnwidth}{!}{
\begin{tabular}{llll}
\hline
                & Femtosecond pulse train & \multicolumn{2}{l}{Picosecond pulse train} \\ \hline
Repetition rate & 15.2 MHz                & 0.3 MHz              & 1.1 MHz             \\ \hline
Pulse duration  & 227 fs                  & 10.31 ps             & 10.31 ps            \\ \hline
Duty cycle      & 3.4$\times$10$^-6$      & 3.1$\times$10$^-6$   & 1.1$\times$10$^-5$  \\ \hline
\multicolumn{4}{c}{Modality 1: TPE fluorescence} \\  \hline
\begin{tabular}[c]{@{}l@{}}Avr. power\\ Pulse energy\\ Peak power\\ \end{tabular} &
  \begin{tabular}[c]{@{}l@{}}0.7 mW\\ 46 pJ\\ 135 W\\ \end{tabular} &
  \begin{tabular}[c]{@{}l@{}}0.7 mW\\ 2.3 nJ\\ 213 W\\ \end{tabular} &
  \begin{tabular}[c]{@{}l@{}}0.7 mW\\ 0.6 nJ\\ 59 W\\ \end{tabular} \\ \hline
  \multicolumn{4}{c}{Modality 2: SHG} \\  \hline
  \begin{tabular}[c]{@{}l@{}}Avr. power\\ Pulse energy\\ Peak power\\  \end{tabular} &
  \begin{tabular}[c]{@{}l@{}}0.3 mW\\ 20 pJ\\ 58 W\\ \end{tabular} &
  \begin{tabular}[c]{@{}l@{}}0.3 mW\\ 1.0 nJ\\ 88 W\\ \end{tabular} &
  \begin{tabular}[c]{@{}l@{}}0.3 mW\\ 0.3 nJ\\ 25 W\\ \end{tabular} \\ \hline
\end{tabular}%
}
\label{tab1}
\vspace{-15pt}
\end{table}

Our result shows that efficient multiphoton imaging is possible using picosecond pulses with the same average power as with the femtosecond pulses; we have shown that reducing \textit{$f_{rep}$} is a very straightforward and effective strategy for improving image quality in two different modalities. Such an approach has several key advantages. Firstly, in the case of chirped pulses typically generated by fiber lasers, it is possible to omit the bulk-optic pulse compressor. This results in a fully-fiberized, maintenance-free portable laser for multiphoton imaging. Secondly, picosecond pulses are significantly less susceptible to chromatic dispersion of microscope optical elements or fiber-based beam delivery systems. This makes the proposed laser system advantageous for multiphoton endoscopic imaging \cite{Jung-15}. We illustrated this issue by measuring the temporal properties of both pulses before and after propagating through the microscope’s optics, with additional dispersive elements, and through numerical simulations. We also investigated the effect on the quality of multiphoton images (discussed in detail in sections 2-4 of the Supplemental Document). Lastly, a significant increase in total fluorescence yield using a lower \textit{$f_{rep}$} (>1 $\mu$s temporal pulse separation) has been reported, enabling dark or triplet state relaxation \cite{Donnert-28}. While generating a low \textit{$f_{rep}$} pulse train directly from the oscillator cavity is possible, the pulse picker approach has the added benefit of flexibility. It allows the user to match experimental conditions, such as scanning speed, signal yield, and sample type.

We also note some disadvantages. Firstly, imaging at a lower \textit{$f_{rep}$} can be a limiting factor for a fast scanning system; at least one pulse per pixel is needed, ultimately limiting pixel dwell time. Still, imaging with a 1 MHz \textit{$f_{rep}$} results in a minimum pixel dwell time of 1 $\mu$s, translating to 3.8 frames per second for a 512$\times$512-pixel image, which is sufficient for most applications. Secondly, this approach requires high pulse energy, which is challenging to produce in fiber lasers without additional amplification. However, all-fiber Yb-doped oscillators can generate pulse energies of >10 nJ.

In conclusion, we have demonstrated the benefits of multiphoton microscopy with a high-energy Yb:fiber NALM oscillator with a reduced \textit{$f_{rep}$}. We have presented that under the same conditions and with a similar starting duty cycle, the 205 fs compressed pulse train resulted in worse images than the longer 10 ps pulse train. This setup is all-fiber and portable, and thanks to the pulse-picking unit, its \textit{$f_{rep}$} can be adjusted according to one’s needs. The approach is immune to the effect of chromatic dispersion, making it easier to use in real-world conditions. These findings open a new way to use multiphoton microscopy, especially appealing for efficient endoscopic applications.

\begin{backmatter}
\bmsection{Funding} National Science Centre, Poland (2021/43/D/ST7/01126)

\bmsection{Acknowledgments} This research was funded in whole by the National Science Centre (NCN) under grant no. 2021/43/D/ST7/01126. For Open Access, the author has applied a CC-BY public copyright license to any Author Accepted Manuscript (AAM) version arising from this submission. We thank Alicja Kwaśny (WUST) for technical assistance during the construction of the multiphoton microscope and Drs Grazyna Palczewska (University of California Irvine, USA), Łukasz Sterczewski, and Jarosław Sotor (WUST) for inspiring discussions.

\bmsection{Disclosures} The authors declare no conflicts of interest.

\bmsection{Data availability} Data underlying the results presented in this paper are available in Dataset 1, Ref. \cite{ref-29}.

\bmsection{Supplemental document} See Supplement 1 for supporting content.

\end{backmatter}

\bibliography{contents}

\bibliographyfullrefs{contents}

\end{document}